\documentclass[a4paper, 12pt, floatfix]{article}
\usepackage[format=hang,singlelinecheck=0,font={sf,small},labelfont=bf]{subfig}
\usepackage{pifont}
\usepackage{graphicx}
\usepackage{epstopdf}
\usepackage[multi-part-units = single]{siunitx}
\usepackage{amsmath}
\usepackage{lineno}
\usepackage{physics}

\usepackage{color}
\usepackage{colortbl}

\newcommand{\TV}[1]{} 
\newcommand{\beq}{\begin{equation}}
	
	\newcommand{\eeq}{\end{equation}}

\newcommand{\e}{\epsilon}
\usepackage[superscript,nomove]{cite}
\usepackage[square,super,comma]{natbib}
\usepackage{float}
\usepackage{authblk}

\graphicspath{{./figures/}}  

\usepackage{comment}

\usepackage{geometry}

\begin{document}



	\title{Observation of light driven band structure via multi-band high harmonic spectroscopy}
	\author[1,2]{Ayelet J. Uzan-Narovlansky}
	\author[3,4]{\'Alvaro Jim\'enez-Gal\'an}
	\author[5]{Gal Orenstein}
	\author[6]{Rui E.F. Silva}
	\author[1]{Talya Arusi-Parpar}
	\author[1]{Sergei Shames}
	\author[1]{Barry D. Bruner}
	\author[7]{Binghai Yan}
	\author[3,8]{Olga Smirnova}
	\author[3,9,10]{Misha Ivanov}
	\author[1]{Nirit Dudovich}
	\affil[1]{\footnotesize Department of Complex Systems, Weizmann Institute of Science, 76100, Rehovot, Israel}
	\affil[2]{\footnotesize Department of Physics, Princeton University, Princeton, NJ, USA}
	\affil[3]{\footnotesize Max-Born-Institut, Max-Born Strasse 2A, D-12489 Berlin, Germany}
    \affil[4]{\footnotesize Joint Attosecond Science Laboratory, National Research Council of Canada and University of Ottawa, Ottawa, Canada}
	\affil[5]{\footnotesize SLAC National Accelerator Laboratory, Stanford University, Stanford, California 94305, USA}
	\affil[6]{\footnotesize Instituto de Ciencia de Materiales de Madrid (ICMM), Consejo Superior de Investigaciones Científicas (CSIC), Madrid, Spain}
	\affil[7]{\footnotesize Department of Condensed Matter, Weizmann Institute of Science, 76100, Rehovot, Israel}
	\affil[8]{\footnotesize Technische Universit\"at Berlin, Ernst-Ruska-Geb\"aude, Hardenbergstr. 36A, D-10623 Berlin, Germany}
	\affil[9]{\footnotesize Blackett Laboratory, Imperial College London, South Kensington Campus, SW7 2AZ London, United Kingdom}
	\affil[10]{\footnotesize Department of Physics, Humboldt University, Newtonstrasse 15, 12489 Berlin, Germany}

	\renewcommand\Authands{ and }
	
	\maketitle 

	\newpage

		\textbf{Intense light-matter interactions have revolutionized our ability to probe and manipulate 
		quantum systems at sub-femtosecond time scales\cite{corkum2007attosecond}, 
		opening routes to all-optical control of electronic currents in solids at petahertz rates\cite{garg2016multi,langer2018lightwave,lucchini2016attosecond,schultze2014attosecond,
			schiffrin2013optical,schultze2013controlling}. Such control typically requires 
		electric field amplitudes  $\sim V/\si{\angstrom}$, when the voltage drop across a lattice site 
		becomes comparable to the characteristic band gap energies. In this regime, intense 
		light-matter interaction induces significant modifications of electronic and optical properties 
		\cite{silva2019topological,jimenez2020lightwave,lakhotia2020laser}, dramatically modifying the crystal band structure. Yet, identifying and characterizing 
		such modifications remains an outstanding problem. As the oscillating electric field changes
		within the driving field's cycle, does the band-structure follow, and 
		how can it be defined? 
		Here we address this fundamental question, 
		proposing all-optical spectroscopy to probe laser-induced closing of the band-gap between 
		adjacent conduction bands. Our work reveals the link between nonlinear light matter interactions in strongly driven crystals and the sub-cycle modifications in their effective band structure.}\\
	The ability of near-resonant light fields to modify the energetics and the dynamics of
	an atom (or a molecule), 
	is central to many phenomena such as laser cooling, trapping, quantum optics 
	of atoms in a cavity \cite{cohen1998nobel,haroche2013nobel}, or
	extremely efficient generation of 
	hyper-Raman lines\cite{sokolov2000raman}. 
	The combined states of matter and near-resonant light are generally studied 
	over many laser cycles using the Floquet formalism, 
	which takes advantage of the periodicity of light
	oscillations. In solids, cycle-averaged modification of the hopping rates between the 
	neighboring sites leads to, e.g., coherent destruction 
	of tunneling \cite{jiangbin2009many} and modified local interaction potentials \cite{lakhotia2020laser}. 
	
	
	The situation changes in light fields with frequencies well below resonance
	but with intensities sufficiently high to induce electron-volts-scale voltages  
	across a lattice site.
	As the oscillating electric field of the lightwave
	changes from zero to its maximum value within a quarter-cycle, 
	rapidly changing voltages can lead to sub-cycle modifications of the macroscopic properties, 
	such as the transmittance\cite{schiffrin2013optical,lucchini2016attosecond} or conductance\cite{schultze2013controlling}. 
	In this regime, a cycle-averaged frequency-domain Floquet perspective 
	is hardly satisfying.   
	
	Here we describe how the effective band structure can be introduced 
	in such interaction regime within a time-domain perspective. We experimentally demonstrate all-optical spectroscopy of a strongly driven crystal, revealing an anomalous spectral intensity response, using high harmonic generation (HHG). Our theoretical study links these observations to laser-induced closing of the gap between adjacent conduction bands (fig. 1a). 
	
	HHG in solids 
	involves optical tunneling\cite{keldysh1965ionization} across the gap between the valence and 
	the conduction band. This transition
	initiates harmonic generation associated with both 
	intraband currents \cite{jurgens2020origin, ghimire2011observation,lakhotia2020laser} and electron-hole recombination,  the latter leading to higher-order 
	harmonic emission \cite{vampa2014theoretical}. 
	At high light intensities and/or low frequencies, one enters a new dynamical regime. In many systems, as the electron-hole 
	wavepacket approaches the edges of the Brillouin zone, 
	Landau-Dykhne-type transitions\cite{hawkins2013role} can 
	promote electrons to higher conduction bands, as reflected in the harmonic spectra\cite{you2017laser,uzan2020attosecond} (fig. 1a). 
	However, can the imprint of these transitions on high harmonic emission  
	be used to observe light-induced modifications of the bands?
	
	One can show that light-induced
	modifications of the bands are directly linked to the sub-cycle Landau-Dykhne-type 
	transitions between them (see Methods). In the low-frequency limit, one  
	begins with the adiabatic approximation, which treats 
	the phase $\omega t$ of the electric field oscillations as a 
	parameter, and finds the adiabatic band structure $\epsilon(k,\omega t)$
	in the presence of a quasistatic electric field. 
	When a non-adiabatic Landau-Dykhne type transition between a pair of bands occurs, 
	typically as the electron is laser-driven across the minimal
	band-gap, it modifies the gap as follows:
	\begin{eqnarray}
		\label{eq:Eq30}
		\Delta \e_{\rm eff}(t)\simeq \Delta \e_{\rm ad}(t)
		\sqrt{\frac{1-w_{\rm LD}(t)}{1+w_{\rm LD}(t)}}
	\end{eqnarray}
	where $\Delta \e_{\rm eff}$ is the effective bandgap, 
	$\Delta \e_{\rm ad}$ is the bandgap in the adiabatic 
	approximation, and $w_{\rm LD}$ is the probability of the sub-cycle 
	Landau-Dykhne type transition. The result is intuitive: 
	the effective band gap is closed by the laser field when the sub-cycle  
	transition approaches unity, $w_{\rm LD}(t)\simeq 1$. Given the
	exponential sensitivity of $w_{\rm LD}(t)$ to 
	laser field strength, the effective bandgap closes soon after the  
	sub-cycle excitation probability becomes appreciable.
	
	To investigate this phenomenon experimentally, we use the two-color HHG spectroscopy\cite{dudovich2006measuring,pedatzur2015attosecond} (fig. 1b), 
	augmenting the strong 
	fundamental driver with
	a weak 
	second harmonic (SH) field, while controlling their subcycle delay $\tau$. Electron-hole trajectories responsible for different harmonics \cite{vampa2015linking}
	are perturbed by the weak field: each trajectory acquires an additional complex phase $\sigma(\tau)$, which is 
	accumulated along the entire trajectory, serving as a sensitive label of its temporal properties. If the fundamental field generates 
	only odd harmonics, the SH field breaks the symmetry of 
	the interaction, as $\sigma(\tau)$ changes sign between the two consecutive half cycles (see SI). 
	This phase is mapped into the harmonic intensity as:
	\beq
	\begin{split}
		I_{odd}(\sigma(\tau))\propto(e^{i\sigma}+e^{-i\sigma})\\
		I_{even}(\sigma(\tau))\propto(e^{i\sigma}-e^{-i\sigma})
		\label{sigma}
	\end{split}
	\eeq 
	Scanning $\tau$ modifies $\sigma(\tau)$ in a periodic manner, modulating the 
	harmonic spectrum. The modulation phase and contrast encode the 
	dynamical properties of  electron trajectories associated with each harmonic order and allow their reconstruction 
	with attosecond precision. In previous studies this method
	resolved the interband contribution to high-harmonic emission\cite{vampa2015linking} and provided
	all-optical reconstruction of the field-free band structure upon exciting  a 
	single conduction band \cite{vampa2015all,vampa2020attosecond}. In this paper we use this scheme to study the underlying dynamics of driven multiband currents, probing the dressed band structure. 

	
	Experiments were performed on MgO\cite{you2017anisotropic}, using $\lambda=1.3 \mu m$ laser field at 
	intensities $\sim10^{13}\text{ }\frac{W}{cm^2}$ and a weak SH field, 
	polarized parallel to the fundamental field. We have 
	measured the HHG modulations with the two-color delay and extracted the oscillation 
	phase $\Phi_N$ associated with each harmonic order $N$. 
	Figure 2 presents $\Phi_N$ as a function of harmonic 
	order for orientation angles of $0^\circ$ ($\Gamma X$) and $45^\circ$ ($\Gamma K$), with respect the fundamental field's polarization. 
	For harmonics $N=11-15$ (10.5-14.5 eV),
	which are associated with the electron-hole recombination 
	from the first conduction band, we measure a gradual slope of $\Phi_N$ 
	with $N$. 
	The slope reflects the evolution of the trajectory length with the harmonic order \cite{vampa2015linking}, supported by our semi-classical calculations (see SI). 
	As we approach the edge of the Brillouin zone, represented by harmonic 17 (16 eV) for $\Gamma K$ and harmonic 19 for $\Gamma X$ directions, this simple description fails.
	A clear phase jump appears in the measurements, associated with the edge of the zone and 
	the changing bands curvature.
	At this point the mapping between momentum and energy becomes singular, leading to the 
	appearance of spectral caustic\cite{uzan2020attosecond}.
	Thus, the phase measurement serves as sensitive probe of the band gap and allows its accurate identification.
	
	Beyond the cut-off energy of the first conduction band, the electron dynamics 
	involve multiple
	bands \cite{schubert2014sub,you2017laser}. The harmonic emission is dictated by electron currents originating from higher conduction bands, determined by their structure and coupling.
	Importantly, these parameters depend on the 
	crystal orientation, with each orientation offering a new one-dimensional 
	slice of the band structure\cite{you2017anisotropic,wu2017orientation}. We can thus 
	track the dependence of electron dynamics on the band structure 
	and laser-induced couplings by resolving the oscillation phase of each harmonic 
	as a function of the crystal orientation. 
	
	Figure 3a presents the HHG oscillation phase, for various crystal orientations.
	The oscillation phases of the lower harmonics (H11,H13,H15) emitted
	far from the edge of the Brillouin zone
	remain unchanged with the crystal orientation. Indeed, the bottom of the first 
	conduction band, as well as its distance from the valence band, $\epsilon_{c1,v}=\epsilon_{c1}-\epsilon_{v}$, is approximately isotropic (fig. 3c). 

	For higher harmonics  ($N=19-29$), a strong variation 
	of the oscillation phase with crystal orientation is observed (fig. 3a). 
	Figure 3b presents the oscillation phase as a function of the crystal's orientation for 
	H21. The phase shows a sharp 
	increase by more than $\pi$ as the orientation angle changes towards 45$^\circ$. 
	The origin of this sharp variation 
	can be understood by looking at the angular dependence of the bands. Figure 3c 
	shows the energy difference between the valence, the first, and the
	second conduction bands ($\epsilon_{c1,v},\epsilon_{c2,v}$) for different crystal orientations, together with the energy contours 
	for harmonic 21. The crystal momentum at the emission point associated with H21 changes quickly with crystal orientation, as illustrated in fig. 3d.
	The strong orientation-dependent modifications of the band gap, $\epsilon_{c2,v}$, lead to significant angular changes of the 
	corresponding electron trajectories, as captured by 
	the phase measurement. In contrast, the 
	emission points for H11-H15 are almost constant with the crystal orientation (fig. 3d). Such nearly isotropic response is captured
	by the phase measurement as well (fig. 3b). 

	We now turn to resolving the light-induced dressing of the band structure. While in large band gap materials, 
	such as MgO, observing band gap modifications requires high intensities, the gap between the conduction bands is relatively small. Therefore, it can be significantly modified at moderate field intensities. We focus on harmonic emission associated with an energy gap between the first and second conduction band. In fig. 4a we plot the field free band structure, which provides a good description of the system at low laser intensities. At $0^\circ$ orientation the minimum energy gap between two conduction bands, $\epsilon_{c2}-\epsilon_{c1}$, is around $3eV$; harmonic 20 is located inside this gap. As we rotate the crystal, the energy gap rapidly reduces to zero, and harmonic 20 is emitted from the second conduction band. In fig. 4b, we plot the oscillation phase of harmonic 20, $\Phi_{20}$, as a function of crystal orientation, for different laser intensities. At low intensity (light green), $\Phi_{20}$ shows a dramatic angular dependence, mainly between $0^\circ-25^\circ$, as a consequence of the strong angular dependence of $\epsilon_{c2}-\epsilon_{c1}$. Importantly, as we increase the field's 
	intensity, the angular dependence of $\Phi_{20}$ decreases and then flattens 
	significantly (dark green). These experimental results are confirmed by our numerical simulations (fig. 4c).
	Numerically, the flattening of the angular dependence of the oscillation phase coincides with
	the onset of strong sub-cycle Landau-Dykhne transitions, with the probability approaching 50\% (see SI).

	
	
	Our theoretical and numerical results link these intensity-dependent observations with
	the closing of the effective band-gap by the laser field. 
	According to equation \eqref{sigma}, the oscillation phase is dictated by the additional complex phase, $\sigma$, induced by the SH field\cite{pedatzur2015attosecond}. The imaginary component, $Im(\sigma)$, is associated with perturbations of the tunneling and recombination probabilities, while the real component $Re(\sigma)$ reflects subtle modifications of the electron trajectory as it propagates within the band. Since the odd and the even harmonics represent constructive and destructive interference of two subcycle emissions, $Re(\sigma)$ leads to an oscillation phase difference. However the imaginary part, $Im(\sigma)$, affects both even and odd harmonics simultaneously, therefore their oscillation phases coincide (see SI).



	In fig. 4d, we plot the relative oscillation phase of harmonics 19 and 20, $\Phi_{20}-\Phi_{19}$, at $0^\circ$ orientation, for different fundamental field intensities. At sufficient low laser intensity, this phase difference vanishes reflecting the dominant role of the imaginary component. The origin of this imaginary component could be associated with an anomalous emission event of these harmonics, emitted from an energy gap. At higher field intensities, the band structure is strongly dressed so that the energy gap between the bands becomes negligible. Therefore, the associated imaginary component is reduced and the perturbation is dominated by the real component $Re(\sigma)$, leading to a phase difference between the even and the odd harmonics (for further discussion see SI)

	In summary, our study establishes all-optical spectroscopy of a strongly driven crystal, revealing laser-induced modification of the band structure. We identify the dynamical transitions between several conduction bands as well as probe their structural dependence. Importantly, we resolve the clear signature of harmonic emission from the energy gap between two conduction bands, probing its modification by the laser field. This study provides a general framework for resolving and interpreting attosecond electronic response phenomena in strongly driven solids. Looking forward, two-color HHG spectroscopy opens a window into the observation a broad range of electronic phenomena -- from sub-cycle phase transitions to ultrafast dynamics in correlated systems -- some of which have been theoretically predicted decades ago, while others are still hotly debated.


	\section*{Acknowledgments}
	N.D. is the incumbent of the Robin Chemers Neustein Professorial Chair. N.D. acknowledges the Minerva Foundation, the Israeli Science Foundation and the European Research Council for financial support. A.J.N.U. acknowledges financial support by the Rothschild Foundation and the Zuckerman Foundation. M.I. acknowledges funding of the DFG QUTIF grant IV152/6-2. Á.J.G. acknowledges funding from the European Union’s Horizon 2020 research and innovation programme under the Marie Skłodowska-Curie grant agreement no. 101028938. A.J.G. and M.I. acknowledge funding from the European Union’s Horizon 2020 research and innovation program under grant agreement No 899794. R. E. F. S. acknowledges support from the fellowship LCF/BQ/PR21/11840008 from “La Caixa” Foundation (ID 100010434) and from the European Union’s Horizon 2020 research and innovation program under the Marie Sklodowska-Curie grant agreement No. 847648.

	\section*{Contributions}
	%
	N.D. and M.I. supervised the study. M.I. supervised the theoretical work and developed the theoretical model. A.J.N.U. and G.O. conceived and planned the experiments. A.J.G. and R.S. performed the theoretical study and the numerical analysis. A.J.N.U., T.A.P., G.O. and B.D.B performed the measurements. A.J.N.U., G.O. and S.S. analyzed the data. B.Y. performed the DFT calculations. All authors discussed the results and contributed to writing the manuscript.
	\section*{Competing financial interests}
	The authors declare no competing financial interests.
	
	\section*{Corresponding authors}
	Correspondence to Nirit Dudovich (nirit.dudovich@weizmann.ac.il).
		\clearpage

	\begin{figure}
		\centering
		\includegraphics[trim= 20 210 20 100,clip,width=1\textwidth]{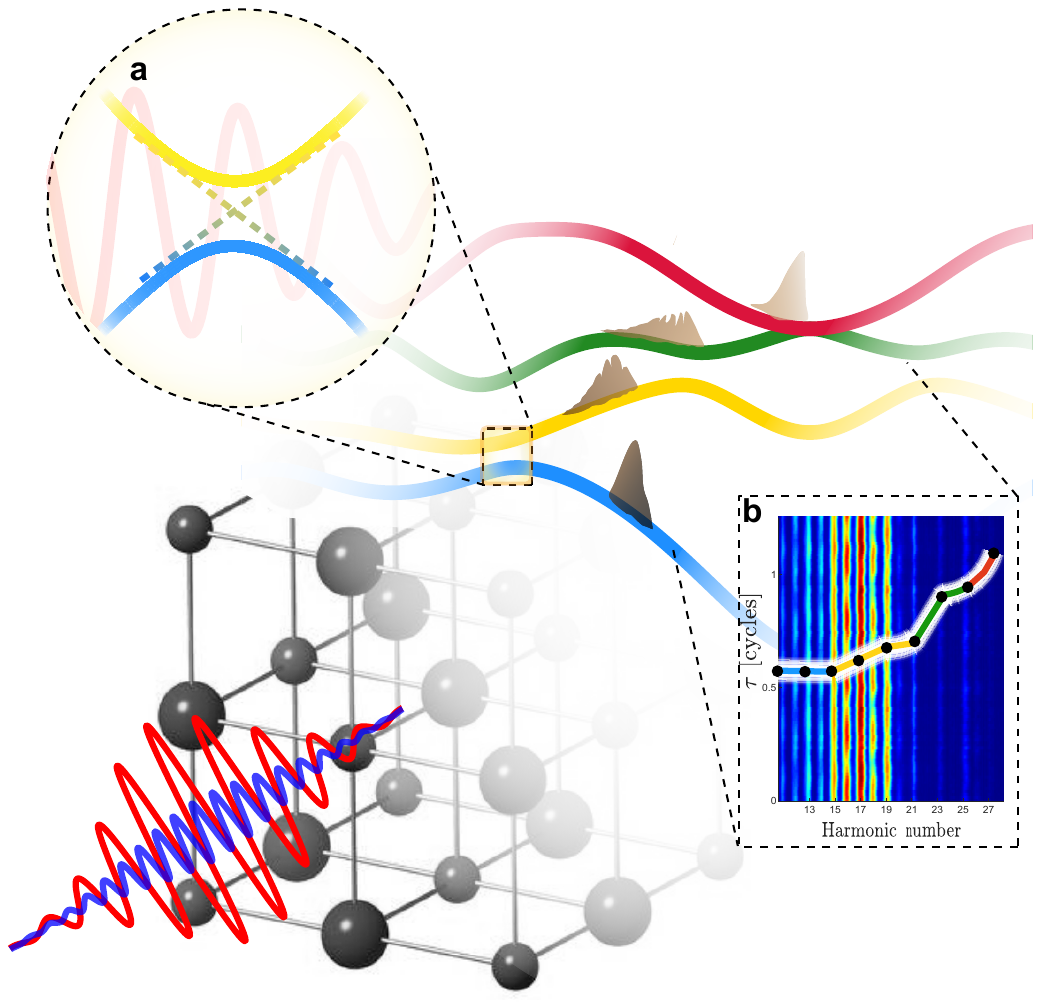}
		\caption{\textbf{All optical spectroscopy of dynamical band structure}. a, Non-adiabatic Landau-Dykhne transition between a pair of bands. b, Two color HHG spectroscopy probes the internal dynamics, mapping the temporal properties of electron trajectories, transitions between the bands as well as their laser driven modifications.}
		\label{fig1:fig1}
	\end{figure}
	\begin{figure}
		\centering
		\includegraphics[trim= 10 280 10 100,clip,width=1.0\textwidth]{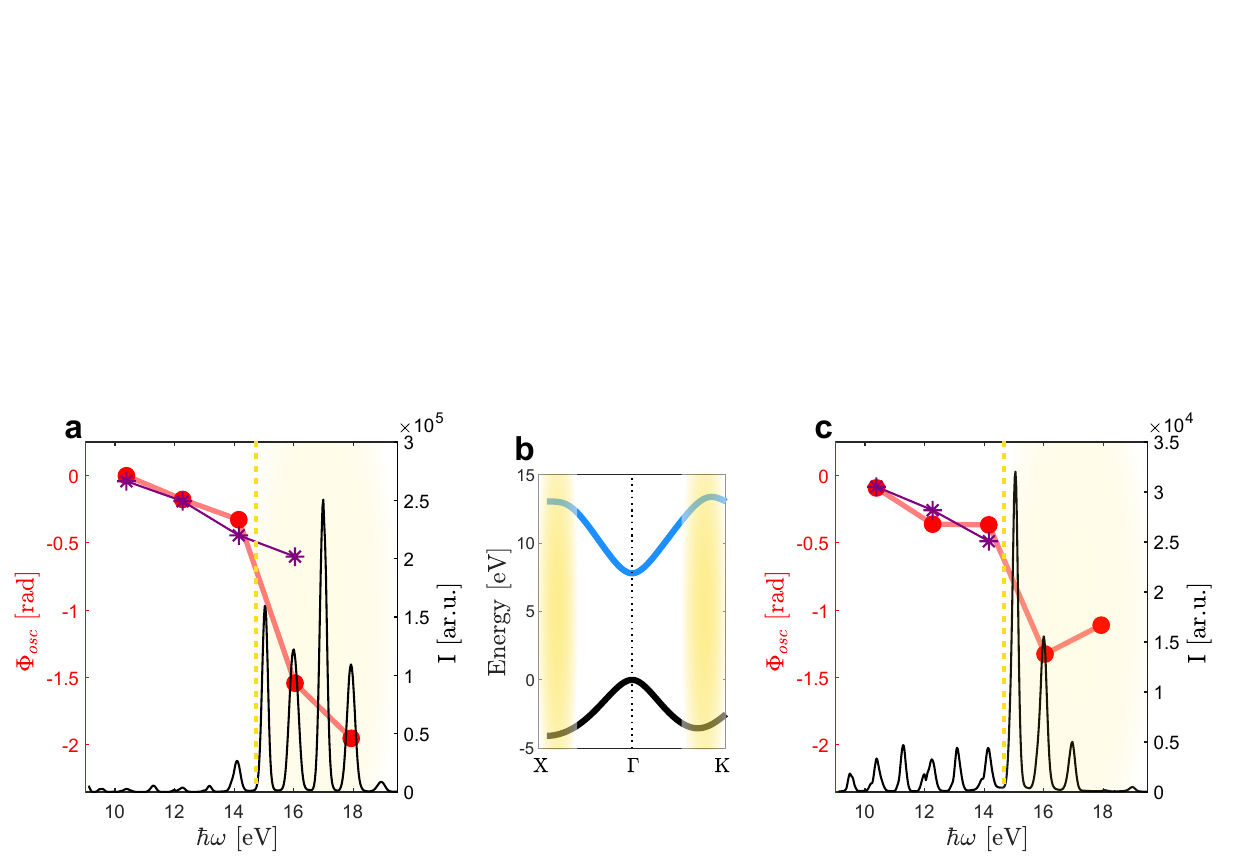}
		\caption{\textbf{HHG spectroscopy beyond the semi-classical description} a,c HHG spectrum (black line) and the oscillation phase (red dots) as function of photon energy for crystals orientations of $0^\circ$(a) and $45^\circ$(c). Calculated oscillation phase using the saddle point approximation in interband model\cite{vampa2014theoretical} (purple star markers). b, The valence and the first conduction band for $0^\circ$ ($\Gamma$ to $X$) and for $45^\circ$ ($\Gamma$ to $K$). The yellow shaded area emphasizes the energy range where the semi-classical description fails\cite{uzan2020attosecond}, also marked at the corresponding photon energies in a and c (dashed yellow line).}
		\label{fig2:fig2}
	\end{figure}
	%

	\begin{figure}
		\centering
		\includegraphics[trim=2cm 4cm 1.8cm 5cm, clip=true, width=1.0\textwidth]{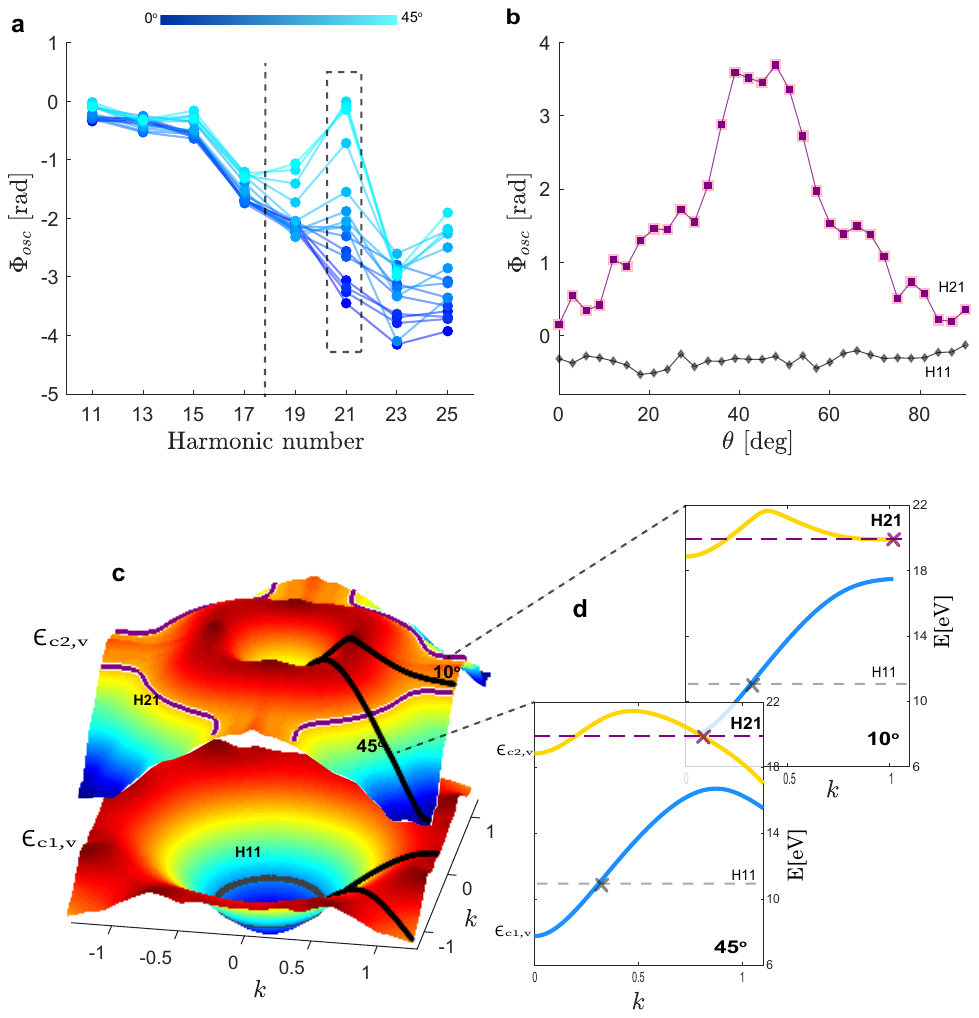}
		\caption{\textbf{Probing the structural dependence of multiple conduction bands} a, The oscillation phase as a function of harmonic number, for different crystal orientation, ranging from $0^\circ$  (dark blue) to $45^\circ$ (cyan) . The dash line marks the cutoff of the first conduction band and the dash box emphasizes H21 oscillation phases for different orientation, plotted at b. b, Harmonic 21 (purple) and harmonic 11 (gray) oscillation phase as a function of crystal orientation. c, top: 2D second band gap, $\epsilon_{c2,v}=\epsilon_{c2}-\epsilon_v$, as function of crystal momentum. The purple contour represents harmonic 21 energy along different crystal's orientation. bottom: 2D first band gap, $\epsilon_{c1,v}=\epsilon_{c1,v}-\epsilon_v$, as function of crystal momentum. The gray contour represents harmonic 11 energy along different crystal's orientation. d, 1D cut of the second (yellow) and first (blue) band gap along $10^\circ$ and $45^\circ$ crystal orientation. The energy of Harmonic 21 and 11 is presented by the purple and gray dash lines as well as their crossing point with the band gaps (cross markers).}
		\label{fig3:fig3}
	\end{figure}
	%

	\begin{figure}
		\centering
		\includegraphics[trim= 50 150 50 150,clip,width=1.0\textwidth]{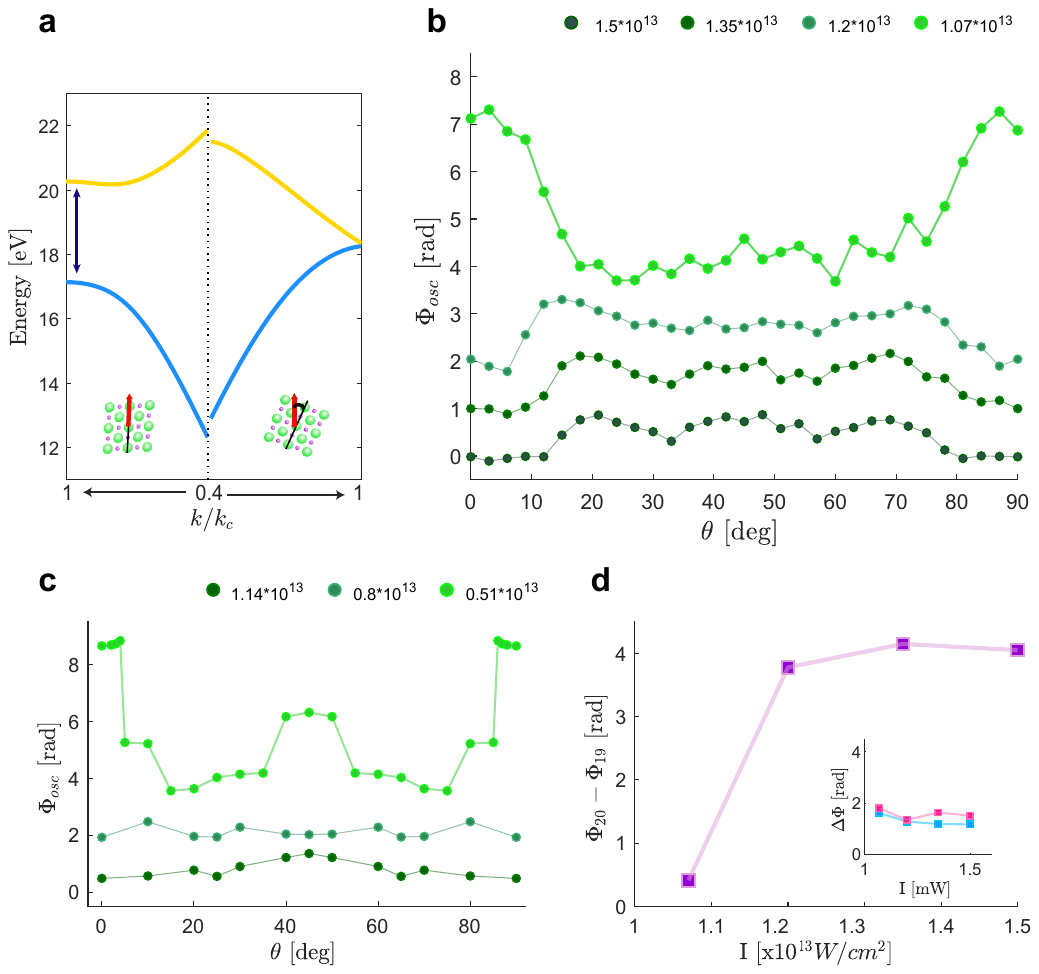}
		\caption{\textbf{Dynamical band structure.} a, First (blue) and second (yellow) band gaps, $\epsilon_{c1,v}$ and $\epsilon_{c2,v}$, for $0^\circ$ (right) and $25^\circ$ (left) orientations, as a function of $k/k_c$ ($k_c=2\pi/a$, where $a$ the lattice constant). b, Harmonic 20 oscillation phase as function of crystal orientation, for different fundamental field intensities (light green to dark green). c, Calculated oscillation phase of H20 as function of the crystal's orientation, for different fundamental field intensities (light green to dark green) d, The oscillation phase difference ,$\Phi_{20}-\Phi_{19}$, at $0^\circ$ orientation as function of the fundamental field intensities. Inset: $\Phi_{18}-\Phi_{17}$ (pink) plot and $\Phi_{22}-\Phi_{21}$ (blue).}
		\label{fig4:fig4}
	\end{figure}
	\clearpage
	\section*{Data availability}
	
The data and datasets that support the plots within this paper and other findings of this study are available from the corresponding author upon reasonable request. The custom code used for the current study has been described in previous publications, and parts of it can be made available from the corresponding author on reasonable request.
	
		\clearpage


	\section*{Methods}
	\subsection*{Dressed Bands in Strongly Driven Solids}
	We consider the response of a two-band solid driven by a strong low-frequency field, and show how
	an effective band structure can be introduced in this case on the sub-cycle time-scale, relevant for the
	low-frequency drivers. In addition to a simple result which averages the instantaneous 
	adiabatic energies over the current wavefunction, we introduce an extension of the Floquet-type
	analysis to the sub-cycle time-scale.
	
	\subsubsection*{Floquet-type analysis on the sub-cycle time-scale: Effective band gap in a strong low frequency field}
	Let us start with some well defined eigenstate, with the wavefunction $\psi_{\kappa}$, which satisfies the stationary Schroedinger equation, $\hat H\psi_{\kappa}=E_g\psi_{\kappa}$. Here $\kappa$ collects the quantum numbers that label the state. In a solid, $\kappa$ labels the crystal momentum \textbf{k} together with the band index $n$.
	
	Let the Hamiltonian depend on some external parameter $\beta$, 
	$\hat H(\beta)$. If we change  $\beta$ slowly, our eigenstate 
	will slowly evolve 
	following the stationary Schroedinger equation, 
	\begin{eqnarray}
		\hat H(\beta)\phi_{\kappa}(\beta)=\epsilon_{\kappa}(\beta) \phi_{\kappa}(\beta)
		\label{eq:Eq1}
	\end{eqnarray}
	
	In the low-frequency field, where the driver has frequency $\omega$
	substantially lower than the energy gap, such parameter $\beta$ is $\beta=\omega t$. 
	From now on we will simply refer to the time $t$.
	In the low-frequency field, the adiabatic states are the solutions of the \textbf{stationary}
	Schroedinger equation with time treated as
	a parameter:
	\begin{eqnarray}
		&&\hat H(t) \phi_{\kappa}(t)=\epsilon_{\kappa}(t)\phi_{\kappa}(t)
		\nonumber \\
		&&\hat H(t)=\hat H_0 + \hat V(t)
		\label{eq:Eq2}
	\end{eqnarray}
	where $V(t)$ is the interaction with the low-frequency  field. 
	The adiabatic eigenstates $\phi_{\kappa}(t)$ form complete basis set.
	Each state has an associated time-dependent state
	$\Psi_{\kappa}$ which incorporates the standard energy phase-factor:
	\begin{equation}
		\Psi_{\kappa}(t)=e^{-i\int_{t_i}^t \epsilon_{\kappa}(t')dt'}\phi_{\kappa}(t) 
		\label{eq:Eq3}
	\end{equation}
	
	Putting this back into the time-dependent Schroedinger
	equation, we see that
	the equation for $\Psi_{\kappa}$ contains extra term proportional to the 
	derivative of the eigenstate:
	\begin{equation}
		i\dot \Psi_{\kappa}=\hat H(t)\Psi_{\kappa} +i e^{-i\int_{t_i}^t \epsilon_{\kappa}(t')dt'}\dot \phi_{\kappa}
		\label{eq:Eq5}
	\end{equation}
	As long as the non-adiabatic Landau-Dykhne transitions between the adiabatic states, caused by this term, are 
	small, their instantaneous energies 
	$\epsilon_{\kappa}(t)$ offer a very good approximation for the effective instantaneous (and hence sub-cycle) 
	energies of the driven system.   
	However, in the presence of strong non-adiabatic transitions between 
	the adiabatic states, these concepts require corrections.
	
	
	Consider now the specific case of a solid, with bands $\epsilon_n(\textbf{k})$ and the Bloch wavefunctions $\phi_{n,\textbf{k}}$, 
	interacting  with a low-frequency laser field. 
	We note that the 
	analysis below is not, in fact, limited to such low-frequency case, but the low-frequency
	case presents the most natural physical situation where our analysis and its conclusions 
	are physically transparent.
	
	With the light-solid interaction treated in the dipole approximation and in the length gauge, 
	the initial crystal momentum $\textbf{k}$ becomes a function of time. We shall label this time-dependent
	momentum
	$\kappa(t)=\textbf{k}+\textbf{A}(t)$. 
	
	It is very useful then to use the Houston states
	to analyze the interaction. In this basis the Bloch wavefunctions $\phi_{n,\textbf{k}}$ 
	and the band energies $\epsilon_n(\textbf{k})$ 
	also follow the vector potential, replacing $\textbf{k}$ with $\kappa(t)=\textbf{k}+\textbf{A}(t)$. The field-free energy $\epsilon_n(\textbf{k})$ goes into  
	$\epsilon_n(\kappa(t))\equiv \epsilon_n(\textbf{k}+\textbf{A}(t))$, so that the band energy ``slides'' 
	with the instantaneous momentum $\kappa(t)$. The Bloch wavefunction
	also ``slides'' with this instantaneous momentum, becoming
	$\phi_{n,\kappa(t)}$. 
	
	In this basis, each time-dependent crystal momentum $\kappa(t)=\textbf{k}+\textbf{A}(t)$ ``traces'' its
	own multi-level system, with its states labeled with the band index $n$ and
	the time-dependent energies and couplings. The multi-level system with crystal momentum $\textbf{k}$
	is decoupled from other multi-level systems with momenta $\textbf{k'}$.

	\subsubsection*{Beyond the adiabatic evolution}

	Consider a two-band solid, with the band indexes $n=1,2$, driven
	by a low-frequency field. Thanks to the small frequency, the adiabatic evolution
	would be a good approximation for the most part of the Brillouin zone.
	The adiabatic evolution will break down to the greatest extent in the regions of the 
	smallest band gap, with exponential dependence on the bandgap. 
	
	Different $\textbf{k}$ will reach these regions at different times
	and hence with different instantaneous values of the field $\textbf{F}(t)$. 
	%
	Suppose we start in the state $\textbf{k}$, turn on the field, and at a moment
	$t_i$ arrive at some momentum $\kappa$ in one of the adiabatic states
	(say with the label $\ket{1,\kappa}$), approaching a region where the two
	bands are coming close to each other. 
	
	We shall now look at the propagator across the region of interest.
	The analysis is general, but for the physical interpretation to be clear, 
	the time-interval should be enough to go through the region. 
	For compactness we will drop the crystal momentum index for the moment.
	
	The propagator, written in the basis of the adiabatic states 
	$\ket{\Psi^{(1)}}, \ket{\Psi^{(2)}}$,
	with energies $\epsilon^{(1)}(t)$ and $\epsilon^{(2)}(t)$, has the following form
	\begin{eqnarray}
		\label{eq:Eq10}
		\hat U(t,t_i)=
		e^{-i\frac{1}{2}\left[\lambda^{(1)}+\lambda^{(2)}\right]} 
		\left(
		\begin{array}{cc}
			\cos\alpha e^{i\lambda} & -\sin \alpha e^{i\phi} \\
			\sin\alpha e^{-i\phi} &  \cos\alpha e^{-i\lambda} \\
		\end{array}
		\right)
		\nonumber
		\\
		\lambda=\frac{1}{2}\left[\lambda^{(2)}-\lambda^{(1)}\right]
	\end{eqnarray}
	
	This form of the propagator is general and meets the key requirements:
	\begin{eqnarray}
		&& 
		|U_{11}|^2+|U_{21}|^2=1
		\nonumber \\
		&& 
		|U_{12}|^2+|U_{22}|^2=1
		\nonumber \\
		&& 
		\hat U\hat U^{\dagger}=\hat U^{\dagger}\hat U=\hat 1
		\nonumber \\
		\label{eq:Eq14} 
	\end{eqnarray}
	The last point also ensures that the wavefunctions remain orthogonal
	during the passage, $\bra{\Psi^{(2)}}\hat U^{\dagger} \hat U\ket{\Psi^{(1)}}=0$.
	
	The meaning of the matrix elements in this propagator is as follows.
	
	\begin{itemize}
		
		\item The phases $\lambda^{(1,2)}(t)$ are associated with the adiabatic
		energies plus, in general, the geometrical Berry phase:
		\begin{eqnarray}
			\lambda^{(i)}(t)=\int_{t_i}^t dt' \epsilon^{(i)}(t')+\gamma_i
			\label{eq:Eq12}
		\end{eqnarray}
		
		The reason we want to complete the passage  across the region of interest is that we want 
		the geometrical phase, associated with this passage, to accumulate fully. 
		But, if we treat the problem fully numerically, then, of course, such requirement is not 
		necessary: the phases $\lambda^{(i)}(t)$ are simply found numerically.
		
		\item The overall factor in front of the propagator sets the zero-energy level as
		\begin{eqnarray}
			\label{eq:Eq11}
			\ev{\epsilon}=\frac{1}{2}\left[\epsilon^{(2)}+\epsilon^{(1)}\right]
		\end{eqnarray}
		through the phases associated with the adiabatic energies.
		
		\item  Irrespective of how the phases $\lambda^{(i)}(t)$ are obtained, their
		physical interpretation remains the same: their time-derivatives have to be associated
		with the adiabatic energies (which are viable and meaningful in the absence of non-adiabatic transitions, i.e.
		when $|\sin\alpha|<< 1$):
		\begin{eqnarray}
			\tilde{\epsilon}^{(i)}(t)=\frac{\partial \lambda^{(i)}(t)}{\partial t}
			\label{eq:Eq13}
		\end{eqnarray}
		The reason to add ``tilde'' above the adiabatic energy is to stress that the adiabatic energies
		$\tilde{\epsilon}^{(i)}(t)$ obtained in such way may not always coincide with the adiabatic energies
		obtained by diagonalizing the adiabatic Hamiltonian, because of
		the presence of the geometric phase. Again, we stress that the phases
		associated with $\tilde \epsilon^{(i)}(t)$ can be extracted numerically. 
		For the physical interpretation  we will need are their 
		derivatives.
		
		\item The off-diagonal elements describe the Landau-Zener-Dykhne non-adiabatic
		transitions between the two adiabatic states. 
		
		\item The phases $\phi$ of these off-diagonal elements
		are determined by the landscape of the bands and the Berry connections (couplings). 
		As we shall see below, $\phi$ will not matter for the effective band structure.
		
		\item The probability of staying in the adiabatic state is $\cos^2\alpha$ and the probability
		of making the transition is $\sin^2\alpha$.
		
	\end{itemize}
	We can now develop the sub-cycle version of the Floquet analysis. To this end, 
	we return to the propagator, dropping the common zero-energy level phase factor for compactness,
	\begin{eqnarray}
		\label{eq:Eq10bis}
		\hat U(t,t_i)=
		\left(
		\begin{array}{cc}
			\cos\alpha e^{i\lambda} & -\sin\alpha e^{i\phi} \\
			\sin\alpha e^{-i\phi} &  \cos\alpha e^{-i\lambda} \\
		\end{array}
		\right)
		\nonumber
		\\
		\lambda=\frac{1}{2}\left[\lambda^{(2)}-\lambda^{1)}\right]
	\end{eqnarray}
	and look for orthogonal states $\ket{\Psi^{\mu}(t_i)}$ with 
	time-dependent quasi-energies $\epsilon^{\mu}(t,t_i)$,  which 
	depend on the crystal momentum $\textbf{k}$. We want these states to 
	behave as if they were the Floquet states for this propagator:
	\begin{eqnarray}
		\label{eq:Eq20}
		\hat U(t,t_i)\ket{\Psi^{\mu}}=e^{i\mu}\ket{\Psi^{\mu}}
		\nonumber \\
		\epsilon^{\mu}=-\frac{\partial \mu}{\partial t}
	\end{eqnarray}
	The positive sign in the phase of the exponent is for convenience 
	because we shall start with the ``lower'' state, which has negative energy.
	The minus sign in the equation for the quasi-energy is related to the positive sign in the exponent
	in the first equation. 
	
	These quasi-energies and the associated states are as close as one can get
	to the effective bands and effective eigenstates in a strongly driven
	system with non-adiabatic transitions. As we shall see below, in the 
	absence of non-adiabatic transitions they, of course, coincide with the
	adiabatic energies and states.
	
	In principle, one can try to find such states for any time-interval after $t_i$, but a
	meaningful time-interval is an interval sufficient 
	to cross the transition region. Once the region is crossed, each
	component of the wavefunction, projected on the adiabatic states, 
	will mostly evolve on the associated adiabatic bands, and in the 
	absence of non-adiabatic transitions these bands are fine and, as mentioned
	above, coincide with the quasi-energies we shall find below.  
	
	The time-dependent ``eigenstates'' of the propagator have two components
	corresponding to the amplitudes in the two adiabatic states $\ket{\phi^{(i)}(t)}$,
	\begin{eqnarray}
		\label{eq:Eq21}
		\ket{\Psi^{\mu}}=
		\left(
		\begin{array}{cc}
			a_{\mu} \\
			b_{\mu}\\
		\end{array}
		\right)
	\end{eqnarray}
	
	The analysis is straightforward. 
	We ask that the determinant of the matrix $\hat U-e^{i\mu}\hat 1$
	is equal to zero, and the solution is found to be
	\begin{eqnarray}
		\label{eq:Eq23}
		\cos\mu=\cos\alpha\cos\lambda
	\end{eqnarray}
	There are two solutions of this equation, $\mu_1=\mu$ and $\mu_2=-\mu$, and the 
	quasienergies $\epsilon^{\mu_1}, \epsilon^{\mu_2}$ are obtained by differentiating $\mu_1$ and $\mu_2$ with respect to time.
	
	The first observation is that, in the absence of non-adiabatic transitions, when
	$\cos\alpha=1$, $\cos\mu=\cos\lambda$, $\mu_1=-\lambda$ and $\mu_2=\lambda$, and the quasi-energies coincide with the adiabatic states. 
	
	Eq.(\ref{eq:Eq23}) is already sufficient to find the effective bandgap, which is
	equal to
	\begin{eqnarray}
		\label{eq:Eq27bis}
		\Delta \epsilon^{\mu}=\left|\epsilon^{\mu_1}-\epsilon^{\mu_2}\right|=2\left|\frac{\partial \mu}{\partial t} \right|         
	\end{eqnarray}
	
	We differentiate the two sides of Eq.(\ref{eq:Eq23}) with respect to
	time and find 
	\begin{eqnarray}
		\label{eq:Eq27}
		\Delta \epsilon^{\mu}=
		\left[
		\tilde \epsilon^{(2)}_{\rm ad}-\tilde \epsilon^{(1)}_{\rm ad}
		\right]
		\left|\cos\alpha\frac{\sin\lambda}{\sin\mu}\right|
	\end{eqnarray}
	where we have used that 
	\begin{eqnarray}
		\label{eq:Eq27bis}
		2\dot \lambda=\tilde\epsilon^{(2)}_{\rm ad}-\tilde \epsilon^{(1)}_{\rm ad}
	\end{eqnarray}
	Using  the relationship $\cos\mu=\cos\alpha\cos\lambda$, we can re-write
	\begin{eqnarray}
		\label{eq:Eq28}
		\left[\cos\alpha\frac{\sin\lambda}{\sin\mu}\right]^2=\frac{\cos^2\alpha\sin^2\lambda}{\cos^2\alpha\sin^2\lambda+\sin^2\alpha}
	\end{eqnarray}
	and hence
	\begin{eqnarray}
		\label{eq:Eq29}
		\Delta \epsilon^{\mu}=
		\left[
		\tilde \epsilon^{(2)}_{\rm ad}-\tilde \epsilon^{(1)}_{\rm ad}
		\right]
		\sqrt{\frac{\cos^2\alpha\sin^2\lambda}{\cos^2\alpha\sin^2\lambda+\sin^2\alpha}}
	\end{eqnarray}
	Finally, we can simplify this expression by taking into account that if the action
	$\lambda(t)$ is large, which is usually the case in the strong, low-frequency field, then
	$\sin^2\lambda(t,t_i)$ is a fast-oscillating function of $t-t_i$. Replacing it with its average, 
	$\sin^2\lambda(t,t_i)=>1/2$,
	and introducing the notation $\sin^2\alpha=w_{\rm LD}$, we
	get the final result for the bandgap between the two quasi-energies
	\begin{eqnarray}
		\label{eq:Eq30}
		\Delta \epsilon^{\mu}=
		\left[
		\tilde \epsilon^{(2)}_{\rm ad}-\tilde \epsilon^{(1)}_{\rm ad}
		\right]
		\sqrt{\frac{1-w_{\rm LD}}{1+w_{\rm LD}}}
	\end{eqnarray}
	This result shows that the band gap collapses when the non-adiabatic 
	Landau-Dykhne transition approaches unity, $w_{\rm LD}\simeq 1$. 
	
	We note that the formalism based on the sub-cycle analogue of the quasi-energy states of 
	a strongly driven system is especially attractive because it naturally merges into the Floquet analysis when
	$t-t_i$ is equal to one period. 
	
	One can also find the quasi-eigenergies in a different way, by using
	only the relationship $\cos\mu=\cos\alpha\cos\lambda$. Namely, one can use this relationship
	to solve the equations for the amplitudes $a^{\mu}$ and $b^{\mu}$, and
	then average the full Hamiltonian over the eigenvectors obtained
	in such way from Eq.(\ref{eq:Eq21}). The 
	result is exactly the same, as expected.


\end{document}